\documentclass[12pt]{article}

\usepackage{amssymb}

\def\bR {{\mathbb{R}}}

\def\bC {{\mathbb{C}}}

\def\R {{\mathbb{R}}}
\def\N {{\mathbb{N}}}
\def\C {{\mathbb{C}}}
 
\def\Z {{\mathbb{Z}}}

\def\Di {\displaystyle}

\def\MR {\mathrm }

\def\sp {\mathrm {sp}}

\def\mM {{\bf M}} 
\def\mX {{\bf X}} 
\def\mh {{\bf h}} 
\def\mg {{\bf g}} 
\def\ess {\mathrm {ess}} 
\def\Carre {\square}

\makeatletter
\@addtoreset{equation}{section}

\makeatother

\newtheorem{theorem}{Theorem}[section]
\newtheorem{lemma}[theorem]{Lemma}

\begin{document}

\bibliographystyle{plain}

\begin{center}
{\Large \bf {Counting function of the embedded eigenvalues 
 for some manifold with cusps, and magnetic Laplacian 
}}
\end{center}

\vskip 0.5cm

\centerline{\bf { Abderemane MORAME$^{1}$
 and  Fran{\c c}oise TRUC$^{2}$}}

{\it {$^{1}$ Universit\'e de Nantes,
Facult\'e des Sciences,  Dpt. Math\'ematiques, \\
UMR 6629 du CNRS, B.P. 99208, 44322 Nantes Cedex 3, (FRANCE), \\
E.Mail: morame@math.univ-nantes.fr}}

{\it {$^{2}$ Universit\'e de Grenoble I, Institut Fourier,\\
            UMR 5582 CNRS-UJF,
            B.P. 74,\\
 38402 St Martin d'H\`eres Cedex, (France), \\
E.Mail: Francoise.Truc@ujf-grenoble.fr }}

\begin{abstract}
We consider a non compact, complete  manifold {\bf{M}} of finite area with cuspidal ends. 
The generic cusp is isomorphic 
to ${\bf{X}}\times ]1,+\infty [$ with metric 
$ds^2=(h+dy^2)/y^{2\delta}.$ 
 {\bf{X}} is a compact manifold  equipped with the metric $h.$ 
 For a one-form $A$ on {\bf{M}} such that in each cusp $A$ is  a non exact one-form 
 on the boundary at infinity, 
we prove that the  magnetic Laplacian  
 $-\Delta_A=(id+A)^\star (id+A)$ satisfies the Weyl asymptotic formula 
 with sharp remainder. We deduce an upper bound 
 for the counting function of the embedded eigenvalues of the Laplace-Beltrami operator  $-\Delta =-\Delta_0 .$  
\end{abstract}

\section{Introduction}

We consider  a smooth, connected $n$-dimensional  Riemannian manifold 
$(\mM, \mg )$, $(n\geq 2), $ such that 
\begin{equation}\label{hyM} 
\mM \; =\; \bigcup_{j=0}^{J}\mM_j  \;  \quad (J\geq 1)\; ,
\end{equation} 
where the $\mM_j$  are open sets of \mM . 
We assume that 
 the closure of $\mM_0$ is compact and that  the other 
$\mM_j$ are cuspidal ends of \mM .  

This means that $\mM_j \cap \mM_k = \emptyset ,$ if $1\leq j<k,$ and 
that  there exists, for any $j,\ 1\leq j\leq J\; ,$ a 
 closed compact 
$(n-1)$-dimensional Riemannian manifold  
$(\mX _j , \mh_j)$ 
 such that $\mM_j\; $ is isometric to 
$\mX _j \times ] a^2_j, +\infty [\; ,\ (a_j>0)$ equipped with the metric 
\begin{equation}\label{gCups} 
 ds_j^2\; =\; y^{-2\delta_j } (\mh_j \; +\; dy^2\ )\; ;\quad (1/n<\delta_j\leq 1). 
\end{equation} 
So there exists a smooth real one-form $A_j \in T^\star (\mX_j) ,$  non exact,  such that 
\begin{equation}\label{hypA} 
 \left \{ 
\begin{array}{ll} i) \  dA_j\neq 0 & \\ 
\MR{or}& \\ 
ii)\  dA_j=0\ \MR{and}\ [A_j]\ \MR{is\ not\ integer .} \end{array} 
\right.
\end{equation} 
In $ii)$ we mean that there exists a smooth closed curve $\gamma$ in $\mX_j$ such 
that 
$$\int_\gamma A_j \; \notin \; 2\pi \Z \; .$$ 
Then one can always find a smooth real one-form $A\; \in \; T^\star (\mM )$ such 
that 
\begin{equation}\label{hypA2} 
\forall \; j,\ 1\leq j\leq J,\quad A\; =\; A_j \quad \MR{on} \ \  \mM_j\; . 
\end{equation} 

We  define the magnetic Laplacian, 
 the Bochner  Laplacian 
\begin{equation}\label{DeA} 
- \Delta_A\; =\; (i\ d + A)^\star (i\ d +A)\; ,
 \end{equation}
 $( i=\sqrt{-1}\; ,\  
 (i\ d+A)u=i\ du+uA\; , \ \forall \; u\; \in \; C^\infty_0(\mM ; \bC ),$ 
 the upper star, $ ^\star ,$ stands for the adjoint between the square-integrable 1-forms and 
 $L^2(\mM )\, )\; ,$ so $d^\star (Z)$ is the usual Hodge-de Rham coodifferential, and \\ 
 $A^\star (Z)=<A;Z>_{T^\star \mM} ,\ \forall Z\in \Lambda_0^1 (\mM ),$ where 
 $\Lambda_0^1 (\mM )$ denotes the vector space of smooth  one-forms 
 with compact support.

As $\mM $ is a complete metric space, by Hopf-Rinow theorem 
$\mM $ is geodesically complete, so 
it is well known, (see \cite{Shu}  ), that $-\Delta_A$ has a unique 
self-adjoint extension on $L^2(\mM )\; ,$ containing in its domain 
$C_0^\infty (\mM ;\bC )\; ,$ the space of 
smooth and compactly supported functions. 
The spectrum of $-\Delta_A$ is gauge invariant~: 
for any $f\in C^1(\mM ; \bR )\; ,\ -\Delta_A$ 
and $-\Delta_{A+df}$ are unitary equivalent, hence they have the same spectrum. 

For a self-adjoint operator $P$ on a Hilbert space $H,$\\  
$$ \sp (P),\ \sp_\ess (P),\ \sp_p (P),\ \sp_d (P)$$ 
will denote 
respectively the spectrum, the essential spectrum, the point spectrum  and the discrete spectrum of $P.$ 
We recall that 
$$\sp (P)=\sp_\ess (P) \cup \sp_d (P),\ \sp_d(P) \subset \sp_p(P)\ \ 
 \MR{and}\  \  \sp_\ess (P) \cap \sp_d (P)=\emptyset .$$ 
\begin{theorem}\label{ThA} 
Under the above assumptions on $\mM $, the 
essential spectrum 
of the Laplace-Beltrami operator  on $\mM ,\ \ -\Delta =-\Delta_0$ is given by 
\begin{equation}\label{Spectess} 
\left \{ \begin{array}{lcl} \sp_\ess (-\Delta) = 
 [0, +\infty [ , & \MR{if} & 1/n<\delta <1\\ 
\sp_\ess (-\Delta )=  [\frac{(n -1)^2}{4}, +\infty [, & \MR{if} & \delta =1 
\end{array} \right . . 
\end{equation}   
($\Di \delta = \min_{1\leq j\leq J} \delta_j \;$)

 When (\ref{hypA}) and (\ref{hypA2}) are satisfied, 
 the  magnetic Laplacian  $-\Delta_A$ has a compact resolvent.
The spectrum $\sp (-\Delta_A) =\sp_d (-\Delta_A) $ is a sequence of non-decreasing 
 eigenvalues $\Di (\lambda_j)_{j\in \N  }\; ,\ \lambda_j\leq \lambda_{j+1}, 
 \ \lim_{j\to +\infty}  \lambda_j=+\infty ,$ such that 
 the sequence of normalized eigenfunctions 
 $(\varphi_j)_{j\in \N }$ is a Hilbert basis of 
$L^2(\mM ).$ Moreover $\lambda_0 >0. \ (\N $ denotes the set of natural numbers).

\end{theorem}

This theorem is not new. The case $A=0$ was proved in \cite{Don2}, 
and the other case in \cite{Go-Mo}, but in the two cases, for   a 
wider class of Riemann metric. We will give a short proof for our simple class 
of Riemann metric, by following the classical method used in \cite{Don1}, 
\cite{Don2} and \cite{Do-Li}.

 For any self-adjoint operator $P$ with compact resolvent, and for 
 any real $\lambda ,\  N(\lambda, P)$ will denote  the number of eigenvalues, 
 (repeated according to their multiplicity),   of 
 $P $ less then   $\lambda ,$ 
\begin{equation}\label{DeNL} 
N(\lambda, P)\; =\; {\rm trace\ }(\chi_{]-\infty ,\lambda [} (P) )\; ,
\end{equation} 
(for any $I\subset \R ,\ \chi_I(x)=1$ if $x\in I$ and $\chi_I(x)=0$ if $x\in \R \setminus I) .$ 
  
The asymptotic behavior of $\; N(\lambda , -\Delta_A )$ 
satisfies the  Weyl formula with the following sharp remainder.

\begin{theorem}\label{ThD} 
Under the above assumptions on $\mM $ and on $A,$ we have the Weyl formula with remainder as $\lambda \to +\infty ,$ 
\begin{equation}\label{De} 
N(\lambda, -\Delta_A)\; =\; |\mM| \frac{\omega_n}{(2\pi )^n}   \lambda^{n/2} 
\; + \; \bf{O}(r(\lambda ) ) \; ,
\end{equation} 
with 
\begin{equation}\label{rL}
 r(\lambda )\; =\; \left \{ \begin{array}{lll} \lambda^{(n-1)/2} \ln (\lambda ), & \MR{if} & 1/(n-1)\leq \delta \\ 
\lambda^{1/(2\delta )} , & \MR{if} & 1/n < \delta < 1/(n-1) 
\end{array} \right . , 
\end{equation} 
$\Di \delta = \min_{1\leq j\leq J} \delta_j \; ,\ 
|\mM |$ is the Riemannian measure of $\mM$  and $\omega_d$ is the euclidian volume of the unit ball 
of $\Di \R^d,\ \omega_d=\frac{\pi^{d/2}}{\Gamma (1+\frac{d}{2})} . $

\end{theorem}
The asymptotic formula (\ref{De}) without remainder is given in \cite{Go-Mo}, 
and with remainder but only for $n=2$ (and $ \delta_j = 1$  for any $ 1\leq j\leq J$) in \cite{Mo-Tr3}.

The Laplace-Beltrami operator  $-\Delta = -\Delta_0$ may have  embedded eigenvalues 
in its essential spectrum  
$\sp_\ess (-\Delta ) .$ 
Let  $N_\ess (\lambda , -\Delta )$ denote the number 
of  eigenvalues of $-\Delta ,$ (counted according to their multiplicity), less then  $\lambda  .$

\begin{theorem}\label{Coro} 
 There exists 
a constant $ C_\mM $ such that, for any  $\lambda >>1,$
\begin{equation}\label{Maj}  N_\ess (\lambda , -\Delta  )\; \leq  \; 
|\mM| \frac{\omega_n}{(2\pi )^n}   \lambda^{n/2} 
\; + \; C_\mM r_0(\lambda ) \;  ,
\end{equation} 
with $r_0(\lambda )$ defined by 
\begin{equation}\label{LrL} 
r_0(\lambda )\; =\; \left \{ \begin{array}{lll}  \lambda^{\frac{n-1}{2}} \ln (\lambda ) , & \MR{if} & 2/n \leq \delta \leq 1\\ 
 \lambda^{\frac{n-(n\delta -1)}{2}} ,& \MR{if} & 1/n < \delta <2/n 
\end{array} \right . \; ; 
\end{equation} 
$\delta $ is the one defined in Theorem \ref{ThD} . 

\end{theorem} 
 
The above upper bound proves that any eigenvalue of $-\Delta $ has finite 
multiplicity. There exist shorter proofs of the multiplicity, 
see for example \cite{Don1} or Lemma B1 in \cite{Go-Mo}.

The estimate (\ref{Maj}) is sharp when $n=2.$ There exist hyperbolic surfaces $\mM $ 
of finite area  so that
$$N_\ess (\lambda , -\Delta  )\; =  \; 
|\mM| \frac{\omega_2}{(2\pi )^2}   \lambda + \Gamma_\mM \lambda^{1/2} \ln (\lambda ) + 
{\bf{O}} (\lambda^{1/2}) \; ,$$ 
for some constant $  \Gamma_\mM .$ See \cite{Mul} for such examples.

Still in the case of surfaces, a compact perturbation of the metric of non compact  
hyperbolic surface $\mM $ 
of finite area can destroy all embedded eigenvalues, see \cite{Col1}. 

For the proof of Theorem \ref{ThD}, we will follow the standard method of partitioning $\mM $ 
and using min-max principle to estimate  the number of eigenvalues by the sum of the ones of the 
Dirichlet operators and Neumann operators associated to the partition. 
In a cusp partition, we will  diagonalize $-\Delta_A$ to an infinite sum of 
Schr\"odinger operators in a half-line, and then we can use standard estimates of the 
number of eigenvalues for those Schr\"odinger operators. 

For the proof of Theorem \ref{Coro}, we will prove that Theorem \ref{ThD} is still valid 
when one changes $A$ into  $\lambda^{-\rho} A,$ for some one-form $A.$ Then  
we will show that the number of embedded eigenvalues  of $-\Delta $
less than $\lambda $ is bounded above 
by the number of  eigenvalues of $-\Delta_{(\lambda^{-\rho} A)} $
less than $\lambda .$

\section{Proofs} 
Since by the Persson \cite{Per} argument used in \cite{Do-Li},  the essential spectrum of an elliptic operator on a manifold is invariant 
by compact perturbation of the manifold, ( see also Prop C3 in \cite{Go-Mo}), we can write 
\begin{equation}\label{spEss} 
\sp_\ess (-\Delta_A)\; =\; \bigcup_{j=1}^{J} \sp_\ess (-\Delta_{A}^{\mM_j,D} )\; ,
\end{equation} 
where $-\Delta_{A}^{\mM_j,D} $ denotes the self-adjoint operator 
on $L^2(\mM_j)$ associated to $-\Delta_A$ with Dirichlet boundary conditions 
on the boundary $\partial \mM_j$ of $\mM_j.$ 

\subsection{Diagonalization of the magnetic Laplacian}

Let us consider a cusp $\mM_j=\mX_j\times ]a_j^2, +\infty [$ equipped with the metric (\ref{gCups}). Then 
for any $u\in C^2(\mM_j),$ 
\begin{equation}\label{laplaMj} 
-\Delta_A u\; =\; -y^{2\delta_j} \Delta_{A_j}^{\mX_j}u\; -\; y^{n\delta_j}\partial_y (y^{(2-n)\delta_j} 
\partial_y u) \; ,
\end{equation} 
where $\Delta_{A_j}^{\mX_j}$ is the magnetic Laplacian on $\mX_j :$ if for  local 
coordinates $\Di \mh_j =\sum_{k, \ell } G_{k \ell }\ dx_k dx_\ell $ and 
$\Di A_j=\sum_{k=1}^{n-1} a_{j,k}\ dx_k,$ then 
$$ 
-\Delta_{A_j}^{\mX_j}\; =\; \frac{1}{\sqrt{\MR{det}(G)}} \sum_{k, \ell } (i\partial_{x_k} + a_{j,k})\left ( 
\sqrt{\MR{det}(G)} G^{k \ell }(i\partial_{x_\ell} +a_{j,\ell})\right ) \; . 
$$ 
We perform the change of variables $y=e^t , $ and define the unitary operator\\
$U :  L^2(\mX_j \times ]2\ln (a_j), +\infty [)\; \to \; L^2(\mM_j)$ , where $\; ]2\ln (a_j), +\infty [$ is equipped 
with the standard euclidian metric  
$dt^2$, by $U(f)=y^{(n\delta_j-1)/2} f.$ Thus 
$L^2(\mM_j)$ is unitarily equivalent to
$L^2(\mX_j \times ]2\ln (a_j), +\infty [),$  and
\begin{equation}\label{newCord} 
-U^\star \Delta_AUf \; = 
\end{equation} 
$$ - e^{2\delta_j t} \Delta_{A_j}^{\mX_j}f   +  
\frac{(n\delta_j -1)[3+\delta_j (n-4)] }{4} e^{2t(\delta_j -1)} f
 -  \partial_t (e^{2t(\delta_j -1)} \partial _t f)\; . $$ 
 \indent  Let us denote by $(\mu_\ell (j) )_{\ell \in \N }$ the increasing sequence of eigenvalues of 
$-\Delta_{A_j}^{\mX_j},$ each eigenvalue repeated according to its multiplicity. \\  
Then 
$-\Delta_{A}^{\mM_j,D} $ is unitarily equivalent 
to $\Di \bigoplus_{\ell =0}^{+\infty} L^{D}_{j,\ell},$ 
\begin{equation}\label{spDirM} 
\sp (-\Delta_{A}^{\mM_j,D})\; =\; \sp ( \bigoplus_{\ell =0}^{+\infty} L^{D}_{j,\ell}) \; ,
\end{equation} 
where $L^{D}_{j,\ell}$ is the Dirichlet operator on 
$L^2 (]2\ln (a_j), +\infty [)$ associated to 
\begin{equation}\label{defLl} 
L_{j,\ell}\; =\;  e^{2\delta_j t} \mu_\ell (j)\; +\; 
\frac{(n\delta_j -1)}{4} [3+\delta_j (n-4)]e^{2t(\delta_j -1)} 
\; -\; \partial_t (e^{2t(\delta_j -1)} \partial _t )\; .
\end{equation} 
The operator  $L^{D}_{j,\ell}$ depends on $A_j$ since  $\mu_\ell (j)$ depends on $A_j$ but we
skip this dependence in notations for the sake of simplicity,
$$ 0\leq \mu_\ell (j) \leq \mu_{\ell +1} (j)\quad \MR{and} \quad 
\lim_{\ell \to \infty} \mu_\ell (j)=+\infty .$$ 
It is well-known that assumption (\ref{hypA}) implies that 
$$0\; <\; \mu_0 (j)\; .$$
As a matter of fact, if $\mu_0 (j)=0$ and $u_0$ is an 
associated eigenfunction, then 
$idu_0=-u_0A_j,$ so $Re(\overline{u_0}du_0)=0,$ and then $|u_0|$ is constant. 
We can assume that $u_0=e^{-i\varphi}$ with $\varphi $ a real 
function. Then locally $d\varphi =A_j$ , which yields
$dA_j=0,$ so for any $\Di x_0\in \mX_j,$ and  for any regular curve 
$\Gamma_{x_0,x}$ joining $x_0$ to $x,$ we have $\varphi(x)=\oint_{\Gamma_{x_0,x}} A_j.$ Therefore $ e^{i\varphi}$ 
will be a well-defined function on  
$\mX_j$ iff part $ii)$ of (\ref{hypA}) is satisfied, (see for example \cite{Hel}).

When $1/n<\delta_j<1,$ another change of variables can be done.
 Precisely we set $y=[(1-\delta_j)t]^{1/(1-\delta_j)},$ and define  
the unitary operator \\ 
$\Di U :\   L^2(\mX_j \times ]\frac{a_j^{2(1-\delta_j)}}{1-\delta_j}, +\infty [)\; \to \; L^2(\mM_j),$ 
 by \    
$\Di  U(f)=y^{(n-1)\delta_j/2} f.$  

Then  we compute  
 $$ -U^\star y^{n\delta_j}\partial _y[y^{(2-n)\delta_j}\partial_yU(f)]\; =$$ 
 $\Di 
-y^{(n+1)\delta_j/2}\partial _y[y^{(3-n)\delta_j/2}\partial_yf]
-\frac{(n-1)\delta_j}{2}y^{2\delta_j -1}\partial_yf 
+\frac{(n-1)\delta_j [(n-3)\delta_j+2]}{4}y^{-2(1-\delta_j)} f,$ \\ 
so using that $y^{\delta_j}\partial_y =\partial_t$ and that 
$t^\rho \partial_t=\partial_t( t^\rho .) -\rho t^{\rho-1},$ 
we get easily  that
\begin{equation}\label{newCordB} 
-U^\star \Delta_A U f = - [(1-\delta_j)t]^{\frac{2\delta_j }{1-\delta_j}} \Delta_{A_j}^{\mX_j}f   +  
\frac{(n-1)\delta_j [(n-3)\delta_j+2]}{4(1-\delta_j)^2t^2} f
 -  \partial^2_t f\; .
\end{equation} 

Thus, in the case  $1/n<\delta_j<1,$ 
equality (\ref{spDirM} )   holds  also when 
 $L^{D}_{j,\ell}$ is the Dirichlet operator on 
$L^2(]\frac{a^{2(1-\delta_j)}}{1-\delta_j}, +\infty [)$ associated to 
\begin{equation}\label{defLlB} 
L_{j,\ell}\; =\;   \mu_\ell (j)  [(1-\delta_j)t]^{\frac{2\delta_j }{1-\delta_j}} \; +\; 
\frac{(n-1)\delta_j [(n-3)\delta_j+2]}{4(1-\delta_j)^2t^2}  
\; -\; \partial^2_t  \; .
\end{equation} 

\subsection{Proof of Theorem \ref{ThA} } 
To study the spectrum, we use the first diagonalization given by  (\ref{spDirM}) and (\ref{defLl}).

If $\mu_\ell (j) >0$ then $\sp (L^{D}_{j,\ell})=\sp_d (L^{D}_{j,\ell})=\{ \mu_{\ell ,k} (j);\ k\in \N \} ,$ 
where $(\mu_{\ell ,k} (j))_{k\in \N }$ is the increasing sequence of eigenvalues of 
$\Di L^{D}_{j,\ell},\ \lim_{k\to +\infty}  \mu_{\ell ,k} (j)=+\infty .$ 

If $\mu_\ell (j)=0$ then $\sp (L^{D}_{j,\ell})=\sp_\ess (L^{D}_{j,\ell})=[\alpha_n,+\infty [ ,$ with 
$\alpha_n=0 $ if $\delta_j<1,$ and $\alpha_n=(n-1)^2/4$ if $\delta_j=1,$ 
\\ 
(by (\ref{defLl}), if $\delta_j=1,\  L^{D}_{j,\ell}u=-\partial_t^2 u+(n-1)^2/4u,$ 
and by (\ref{defLl}), if $1/n<\delta_j<1,\ 
L^{D}_{j,\ell}u=-\partial_t^2 u+V(t)u$ with $\Di \lim_{t\to \infty} V(t)=0).$

Since we have $\mu_0(j)=0$ when $A=0,$  we get that $\sp_\ess (-\Delta_0)=[\alpha_n,+\infty[ .$

If $A$ satisfies assumptions (\ref{hypA}) and (\ref{hypA2}), 
we have seen that $0<\mu_0(j),$ 
then $0< \mu_\ell (j)$ for all 
$j$ and $\ell ,$ and then 
$$\sp (-\Delta_{A}^{\mM_j,D})=\{ \mu_{\ell ,k} (j) ;\ (\ell ,k)\in \N^2 \} .$$ 
As $\Di \mu_{\ell} (j) \leq \mu_{\ell ,k} (j)< \mu_{\ell ,k+1} (j)$ with 
$\Di \lim_{\ell \to +\infty } \mu_{\ell} (j)=+\infty $ 
and $\Di  
\lim_{k \to +\infty } \mu_{\ell ,k} (j)= +\infty ,$ each $\mu_{\ell ,k} (j)$ 
is an eigenvalue of $-\Delta_{A}^{\mM_j,D}$ of finite multiplicity, so 
$\sp (-\Delta_{A}^{\mM_j,D})=\sp_d (-\Delta_{A}^{\mM_j,D}).$ 
Therefore, we get that 
$\sp_\ess (-\Delta_A)=\emptyset \; \Carre $

\subsection{ Proof of Theorem \ref{ThD} } 
 We proceed as in \cite{Mo-Tr3}. 

We begin by establishing  for $\mM_j$, $1\leq j\leq J$, formula (\ref{De}) with $-\Delta_{A}^{\mM_j,D}$ defined in
 (\ref{spEss}) instead of 
$-\Delta_{A}.$ 
When 
$\delta_j=1$  we use the decomposition given by  (\ref{spDirM}) and (\ref{defLl}), but when  $1/n<\delta_j<1,$ we use the decomposition given by  (\ref{spDirM}) and (\ref{defLlB}).

From now on, any constant depending only on  $\delta_j$ and on 
  $\displaystyle \min_{j }\mu_0 (j) $  will be  invariably denoted by
 $C\; .$ 

As in \cite{Mo-Tr3}, we will follow Titchmarsh's method. Using Theorem 7.4 in  \cite{Tit} page 146,  
we prove the following Lemma.

\begin{lemma}\label{Titch} 
There exists  $C > 1 $ so that for any 
$\lambda   >> 1$ \\ 
and any $\Di \ell \in K_\lambda =\{l\in\N ; \  \mu_\ell (j)\in  [0, \lambda/\min_{j} a_j^{4\delta_j} [  \}\; ,$ 
\begin{equation}\label{tit}  
| N(\lambda , L^{D}_{j,\ell}) \; -\;  \frac{1}{\pi} 
w_{j,\ell}  (\lambda  ) | 
\; \leq \; 
 C \ln ( \lambda )
 \; ,
 \end{equation} 
with 
$\Di \ 
w_{j,\ell} (\mu )\; =\; 
\int_{\alpha_j }^{+\infty}  \left [ 
\mu   -  V_{j,\ell}  (t) \right ]_{+}^{1/2} dt  
\; =\; 
\int_{\alpha_j }^{T_j (\mu )}  \left [ 
\mu   -  V_{j,\ell}  (t) \right ]_{+}^{1/2} dt . 
$ \\
The potential $ V_{j,\ell} $ is defined as following: 
\begin{equation}\label{defwll} 
\left \{ \begin{array}{l} \MR{if}\  \delta_j=1 \\
V_{j,\ell}  (t)= \mu_\ell (j )e^{2t} +\frac{(n-1)^2}{4} \\ 
\MR{if}\ 1/n <\delta_j <1\\ 
V_{j,\ell}  (t)= \mu_\ell (j) [ (1-\delta_j)t]^{\frac{2\delta_j }{1-\delta_j}} \; +\; 
\frac{(n-1)\delta_j [(n-3)\delta_j+2]}{4(1-\delta_j)^2} t^{-2}   
\end{array} \right .  , 
\end{equation}
and \begin{equation}\label{defwlls} 
\left \{ \begin{array}{l} \MR{if}\  \delta_j=1 \\ 
\alpha_j = 2\ln (a_j),\quad \Di T_j (\mu )=\frac{1}{2} \ln \left(\mu /\mu_0(j)\right)\\ 
\MR{if}\ 1/n <\delta_j <1\\ 
\alpha_j = \frac{ a_j^{2(1-\delta_j)}}{1-\delta_j},\quad \Di T_j (\mu )=\frac{1}{1-\delta_j} \left(\frac{\mu}{\mu_0(j)}\right)^{\frac{1-\delta_j}{2\delta_j}} 
\end{array} \right .  .
\end{equation}

\end{lemma} 

\noindent 
{\bf Proof of Lemma \ref{Titch} }

When $1/n < \delta_j <1,$ by enlarging $\mM_0$   and reducing $\mM_j,$ we can take $\alpha_j$  large enough so that 
$V_{j,\ell}  (t)$ is an increasing function on $[\alpha_j, +\infty [$ 
and $\lambda /\mu_\ell (j) >>1$ when $\ell \in K_\lambda .$ 
 Then, if $\alpha_j\leq Y<X(\lambda )=V^{-1} _{j,\ell}  (\lambda),$ 
  following the proof of Theorem 7.4 in  \cite{Tit} pages 146-147, 
  we get that 
  \begin{equation}\label{TitchEst} 
  | N(\lambda , L^{D}_{j,\ell}) - \frac{1}{\pi}  w_{j,\ell} (\lambda  ) | 
  \leq 
  \end{equation} 
$$C [\ln (\lambda - V_{j,\ell}  (\alpha_j)) - \ln (\lambda - V_{j,\ell}  (Y)) 
  + (X(\lambda ) -Y)(\lambda -V_{j,\ell}  (Y) )+ 1] .$$ 
  When $\delta_j=1,$ we choose $Y=X(\lambda ) -\frac{\sqrt{\ln \lambda }}{\sqrt{\lambda}}. $ \\ 
   When $1/n <\delta_j <  1,$ we choose $\Di Y=X(\lambda ) -\frac{\sqrt{ \ln \lambda }}{\sqrt{\lambda}} 
  \left ( \frac{\lambda} {\mu_\ell (j)} \right )^{\frac{1-\delta_j}{4\delta_j}} ; $ 

$
  ( X(\lambda ) \sim \frac{1}{1-\delta_j}\left ( \frac{\lambda} {\mu_\ell (j)} \right )^{\frac{1-\delta_j}{2\delta_j}} ) \; \Carre   $ 

 Let us apply to $-\Delta_{A_j}^{\mX_j}$ , the magnetic Laplacian which lies on $\mX_j,$ on  a "boundary at infinity", the sharp asymptotic Weyl formula of L. H\"ormander \cite{Hor1} (see also \cite{Hor2}), 
 
\begin{theorem}\label{Horm} 
There exists  $C > 0 $ so that for any 
$\mu    >> 1\; $ 
\begin{equation}\label{WeylH} 
|N(\mu , -\Delta_{A_j}^{\mX_j}) \; - \; \frac{\omega_{n-1}}{(2\pi )^{n-1}} |\mX_j | \mu^{(n-1)/2} |\; \leq \; 
C \mu^{(n-2)/2}\; . 
\end{equation} 
\end{theorem}

\begin{lemma}\label{DirCusp} 
There exists  $C > 0 $ such  that for any 
$\lambda   >> 1\; $ 
\begin{equation}\label{NLcuspD}
|N(\lambda , -\Delta_{A}^{\mM_j,D}) -\frac{\omega_n}{(2\pi )^n} |\mM_j| \lambda^{n/2} | 
\; \leq 
\end{equation} 
$$ C  \left \{ \begin{array}{lll} \lambda^{(n-1)/2} \ln (\lambda ) , & \MR{if} & 1/(n-1)\leq \delta_j \leq 1\\ 
\lambda^{1/(2\delta_j)} , & \MR{if} & 1/n < \delta_j < 1/(n-1)
\end{array} \right . 
\; . $$
\end{lemma} 

\noindent 
{\bf Proof of Lemma \ref{DirCusp} } 
By  formula (\ref{spDirM}), 
\begin{equation}\label{NL1}
N(\lambda , -\Delta_{A}^{\mM_j,D})\; =\; \sum_{\ell =0}^{+\infty} 
N(\lambda , L^{D}_{j,\ell}) \; . 
\end{equation}  
 When $\ell  \; \notin \;  K_\lambda , 
\ (K_\lambda$ 
 is defined in Lemma \ref{Titch}), and thanks to formula (\ref{defwll}) we have $ V_{j,\ell} \geq \mu_\ell(j) a_j^{4\delta_j}\geq\lambda$
so  $N(\lambda , L^{D}_{j,\ell})=0$. 
Therefore the estimates (\ref{tit}), (\ref{WeylH}) and formula (\ref{NL1}) 
prove that 
\begin{equation}\label{NL2} 
| N(\lambda , -\Delta_{A}^{\mM_j,D})\;  - 
\sum_{\ell =0}^{+\infty}  \frac{1}{\pi} 
w_{j,\ell}  (\lambda  )  | 
\; \leq \; C \lambda^{(n-1)/2} \ln (\lambda )\; . 
\end{equation} 
Let us denote 
\begin{equation}\label{NL3} 
\Theta_j (\lambda )\; =\; \sum_{\ell =0}^{+\infty}  \frac{1}{\pi} 
w_{j,\ell}  (\lambda  )  \quad \MR{and} \quad R_j(\mu  )\; =\; \sum_{\ell =0}^{+\infty} [\mu - \mu_\ell (j)]_{+}^{1/2} \; .
\end{equation} 
As $\Di R_j(\mu  )=\frac{1}{2} \int_{0}^{+\infty} [\mu -s]_{+}^{-1/2} N(s, -\Delta_{A_j}^{\mX_j} ) ds,$\\ 
the H\"ormander estimate (\ref{WeylH}) entails the following one. 

There exists a constant $C>0$ such that, for any $\mu >>1,$ 
\begin{equation}\label{WeilR}  
| R_j(\mu  )\; -\; \frac{\omega_{n-1}}{2(2\pi )^{n-1}}  |\mX_j | 
\int_{0}^{+\infty} [\mu -s]_{+}^{-1/2} s^{(n-1)/2} ds |\; \leq \; C\mu^{(n-1)/2} \; . 
\end{equation} 

Writing in (\ref{defwll} ) 
\begin{equation}\label{defVj} 
V_{j,\ell } (t)=\mu_\ell (j) V_j(t) \; +\;  W_j(t) \; ,
\end{equation} 
we get that $\Di \Theta_j (\lambda )\; =\;  \frac{1}{\pi}  
\int_{\alpha_j}^{T_j(\lambda )} V_{j}^{1/2}(t) R_j(\frac{\lambda -W_j(t)}{V_j(t)} )dt\; . $\\ 
So according to (\ref{WeilR})  
\begin{equation}\label{estW} 
|  \Theta_j (\lambda )\; -\; 
\frac{\omega_{n-1}\Gamma (\frac{1}{2})\Gamma (\frac{n+1}{2})}{(2\pi )^{n}\Gamma (1+\frac{n}{2})}  
  |\mX_j |  \int_{\alpha_j}^{T_j(\lambda )} 
\frac{(\lambda -W_j(t))^{n/2}}{V_{j}^{(n-1)/2}(t)} dt  |\; \leq \; 
\end{equation} 
$$ C
\int_{\alpha_j}^{T_j(\lambda )}  \frac{(\lambda -W_j(t))^{(n-1)/2}}{V_{j}^{(n-2)/2}(t)}  dt  \; .
$$ 
\indent 
 From  the definitions (\ref{defwll}) and (\ref{defVj}) we get that 
 \begin{equation}\label{estIW} 
|  \int_{\alpha_j}^{T_j(\lambda )} 
\frac{(\lambda -W_j(t))^{n/2}}{V_{j}^{(n-1)/2}(t)} dt \; -\; \lambda^{n/2} \frac{1}{(\delta_j n -1)a_{j}^{2(\delta_j n -1)}} 
|\; \leq \; C \lambda^{(n-1)/2} \; , 
\end{equation}  
and 
 \begin{equation}\label{estIR}  
 \int_{\alpha_j}^{T_j(\lambda )}  \frac{(\lambda -W_j(t))^{(n-1)/2}}{V_{j}^{(n-2)/2}(t)}  dt  \; 
 \leq \; 
   \end{equation} 
$$ C \left \{ \begin{array}{lll} \lambda ^{(n-1)/2} & \MR{if} & 1/(n-1) < \delta_j \leq 1 \\ 
  \lambda ^{(n-1)/2} \ln \lambda  & \MR{if} & 1/(n-1) = \delta_j  \\ 
   \lambda ^{1/(2\delta_j)} & \MR{if} & 1/n < \delta \leq 1/(n-1)  
   \end{array} \right . . 
$$ 
 \indent   
 As $\Di |\mM_j| =\frac{|\mX_j|
}{(\delta_j n -1)a_j^{2(\delta_j n -1)}},$ we get (\ref{NLcuspD}) from (\ref{NL2}), 
 (\ref{NL3}) and  (\ref{estW})--- (\ref{estIR}) $\; \Carre $  

To achieve the proof of Theorem \ref{ThD},  
we proceed as in 
\cite{Mo-Tr3}.

We denote 
$\Di \; \mM_0^0=\mM \setminus (\bigcup_{j=1}^{J}\overline{\mM_j} )\; ,$ 
then 
\begin{equation}\label{partit} 
\mM \; =\; \overline{\mM_0^0}\bigcup  \left (\bigcup_{j=1}^{J}\overline{\mM_j}\right ) \ .
\end{equation} 
Let us denote respectively by $-\Delta_{A}^{\Omega ,D}$ and by $-\Delta_{A}^{\Omega ,N}$ the Dirichlet 
operator and the Neumann-like operator on an open set 
$\Omega \; $ of $\mM $  
associated to  $-\Delta_{A}\; . $ \\ 
$ -\Delta_{A}^{\Omega ,N}$ is the Friedrichs extension defined by 
the associated quadratic form $\Di q_A^\Omega (u)=\int_\Omega |i du+Au|^2d{\bf{m}}\; ,\ u\in 
C^\infty (\overline{\Omega}; \C ),\ u$ with compact support 
in $\overline{\Omega} . \ (d{\bf{m}}$ is the $n$-form volume of 
$\mM $ and $|Z|^2=<Z;Z>_{T^\star (M)}$ for any complex 
one-form $Z$ on $\mM ).$   \\  
The min-max principle and (\ref{partit}) imply that
 \begin{equation}\label{Delll} 
N(\lambda, -\Delta_{A}^{\mM_0^0 ,D}) + \sum _{1\leq j\leq J} N(\lambda, -\Delta_{A}^{\mM_j ,D}) \leq N(\lambda, -\Delta_{A})  
\end{equation}
$$\leq N(\lambda, -\Delta_{A}^{\mM_0^0 ,N}) + \sum _{1\leq j\leq J} N(\lambda, -\Delta_{A}^{\mM_j ,N})$$
\indent 
The Weyl formula with remainder, (see  \cite{Hor2} for Dirichlet boundary condition  and 
\cite{Sa-Va} p. 9 for Neumann-like boundary condition), gives that 
\begin{equation}\label{WM0}  
 N(\lambda , -\Delta_{A}^{\mM_0^0 ,Z})= \frac{\omega_n}{(2\pi )^n}  |\mM_0^0| 
 \lambda^{n/2}  + {\bf{O}}(  \lambda^{(n-1)/2}) \; ; 
\quad (\MR{for}\ Z=D\ \MR{and\ for}\ Z=N) 
 \; .\end{equation} 
For $1\leq j\leq J,$ the asymptotic formula for $ N(\lambda, -\Delta_{A}^{\mM_j ,N})\; ,$ 
\begin{equation}\label{Neumann} N(\lambda, -\Delta_{A}^{\mM_j ,N})\; =\; 
\frac{\omega_n}{(2\pi )^n}  |\mM_j| 
 \lambda^{n/2}  + {\bf{O}}(  r(\lambda ))  
\; , 
\end{equation} 
is  obtained as for the Dirichlet case  
(\ref{NLcuspD})  by 
noticing that   
$$ N(\lambda, L^{D}_{j,\ell }) \leq  N(\lambda, L^{N}_{j,\ell }) 
\leq N(\lambda, L^{D}_{j,\ell }) +1\; ,$$ 
 where $  L^{D}_{j,\ell }$ 
and $L^{N}_{j,\ell }$ are Dirichlet and Neumann-like  
operators on a half-line $I=]\alpha_j , +\infty [\; ,$ 
associated to the same differential Schr\"odinger 
operator 
$\Di L_{j,\ell }$ 
 defined by (\ref{defLl}) when $\delta_j=1,$ and by (\ref{defLlB}) otherwise. \\ 
 (The Neumann-like boundary condition is of the form  
$\partial_tu(\alpha_j) +\beta_j u(\alpha_j)=0$ because of the change of functions performed by $U^\star  ) . $ \\ 
The above inequality is well-known. It comes from the fact that the eigenvalues of 
$L^{D}_{j,\ell }$ and $L^{N}_{j,\ell }$ are of multiplicity one 
and there is no common eigenvalue, (we have used Theorem 2.1. page 225 of \cite{Co-Le}). 
If $(\mu_{\ell ,k }^{Z}(j))_{k\in \N} $ is the sequence 
of non-decreasing eigenvalues of $L^{\overline{Z}}_{j,\ell },\ (Z=D$ or $D=N),$ and $(\varphi_{\ell ,k }^{Z})_{k\in \N} $ an associated 
orthonormal basis of eigenfunctions,    then 
$\mu_{\ell ,0 }^{N}(j)< \mu_{\ell ,0 }^{D}(j).$ 
As  in  $E_{k+1}(Z),$ the subspace of dimension $k+1$ spanned by 
$ \varphi_{\ell ,0 }^{Z},\ \varphi_{\ell ,1 }^{Z}, \ldots , \varphi_{\ell ,k }^{Z},$ 
 there exists, in $E_{k+1}(Z),$ a subspace of dimension 
$k$ included in the domain of $L^{\overline{Z}}_{j,\ell },$ for 
$(Z, \overline{Z})=(N,D)$ and for $(Z,\overline{Z})=(D,N),$ 
the min-max principle involves 
$\Di \mu_{\ell ,k-1 }^{\overline{Z}}(j) < \mu_{\ell ,k }^{Z}(j).$  
(For any $\Di k, \ \varphi_{\ell ,k+1 }^{N} -\frac{\varphi_{\ell ,k+1 }^{N}(\alpha_j)}{\varphi_{\ell ,k }^{N}(\alpha_j)} 
\varphi_{\ell ,k}^{N}$ is in the domain of $L^{D}_{j,\ell }$ 
and $\Di \varphi_{\ell ,k+1 }^{D} -\frac{\partial_t \varphi_{\ell ,k+1 }^{D}(\alpha_j)}{\partial_t \varphi_{\ell ,k }^{D}(\alpha_j)} 
\varphi_{\ell ,k}^{D}$ is in the domain of $L^{N}_{j,\ell }).$

We get (\ref{De})  
from (\ref{NLcuspD}) and  (\ref{Delll})--- (\ref{Neumann}) $\; \Carre $


\subsection{Proof of Theorem \ref{Coro}}

\begin{lemma}\label{FirstEige} 

For any $j\in \{ 1, \ldots ,J\} ,$ there exists a one-form 
$A_j$ satisfying (\ref{hypA})  and the following property. 

There exists $\tau_0 =\tau_0(A_j)>0$ and $C=C(A_j)>0$  such that,  if $\Di \mu_0(j,\tau  )= \inf_{u\in C_0^\infty (\mX_j),\ \| u\|_{L^2(\mX_j)} =1} 
\| idu+\tau uA_j \|^{2}_{L^2(\mX_j)}$ 
denotes the first eigenvalue of $-\Delta_{\tau A_j}^{\mX_j} ,$ then 
\begin{equation}\label{FirstEigeEs} 
\mu_0(j,\tau  )\; \geq \; C\tau ^2\; ;\quad \forall \; \tau \in ]0,\tau_0]\; .
\end{equation} 
($\| idu+\tau uA_j \|^{2}_{L^2(\mX_j)}=\int_{\mX_j} <idu+\tau uA_j; idu+\tau uA_j>_{T^\star (\mX_j)} d{\bf{x}}_j \; ) .$ 
\end{lemma} 
{\bf {Proof of Lemma \ref{FirstEige}. }} 
When $n=2,$ we can take $A_j=\omega_jd{\bf{x}}_j ,\ (d{\bf{x}}_j$ is the $(n-1)$-form volume of $\mX_j ), $ 
for some constant $\Di \omega_j \in \R \setminus \frac{2\pi}{|\mX_j|} \Z ,$ 
then $\mu_0(j,\tau  )=\tau^2 \omega_j^2$ for small $|\tau |.$

When  $n\geq 3, $ 
 we have $\mu_0(j,0 )=0,\ \partial_\tau \mu_0(j,0 )=0$ and 
$$\partial^2_\tau \mu_0(j,0  )\; =\; \frac{2}{|\mX_j|}\int_{\mX_j} \left [ \;  |A_j|^2 -
(-\Delta_{0}^{\mX_j} )^{-1} (d^\star A_j).(d^\star A_j)\; \right ] d{\bf {x}}_j \; .$$ 
 $(d^\star $ is the Hodge-de Rham codifferential on $\mX_j,$  and  $(-\Delta_{0}^{\mX_j} )^{-1} $ 
is the  inverse of the Laplace-Beltrami operator on functions, which is well-defined on the orthogonal of the first eigenspace, 
on the space
 $\{ f\in L^2(\mX_j);\ \int_{\mX_j} fd{\bf{x}}_j = 0\} ).$ 

 The proof is standard. One writes 
 $-\Delta_{\tau A}^{\mX_j} =P_0+\tau P_1+\tau^2P_2,$\\ 
 $P_0=-\Delta_0$ and for all $u\in C^1(\mX_j),\ 
 P_1(u)=i<du;A_j>_{T^\star \mX_j} -id^\star (uA_j)$ and  
 $ P_2(u)=u|A_j|^2=u<A_j;A_j>_{T^\star \mX_j}. $
    The first eigenvalue of $P_0, \ \mu_0(j,0)=0$ is 
 of multiplicity one. The associated normalized eigenfuction is $u_0=1/\sqrt{|\mX_j|} .$ 
 Then $\tau \to \mu_0(j,\tau )$ is an analytic function, and there exists an associated eigenfunction $u_{0,\tau }$ analytic in $\tau . $ 
 Then, as $\tau \to 0,\  \mu_0(j,\tau  )=\tau c_1+\tau^2c_2 +{\bf{O}}(\tau^3)$ 
 and $u_{0,\tau }=u_0+\tau v_1 +\tau^2 v_2 
 +{\bf{O}}(\tau^3),$ with 
 $$\left \{ \begin{array}{l} c_1=\int_{\mX_j} P_1(u_0).\overline{u_0} d{\bf{x}}_j\\ 
 v_1=-P_{0}^{-1}[P_1(u_0)-c_1u_0]\\ 
 c_2=\int_{\mX_j} [P_2(u_0)+P_1(v_1)].\overline{u_0} d{\bf{x}}_j \; .
  \end{array} \right . $$ 
 The operator   $ P_1$ 
 is formally self-adjoint 
 and $P_2$ is self-adjoint. \\ 
 We have  $\Di P_1(u_0)=-\frac{i}{\sqrt{|\mX_j|}} d^\star (A_j)$ so $P_1(u_0)$
 is orthogonal to the constant function  $u_0$ and then $  c_1=0.$  

To the  non-negative quadratic form $A_j\; \to \; \partial^2_\tau \mu_0(j,0  ),$ we associate 
a self-adjoint operator $P$ on $\Di \Lambda^1 ( \mX_j),\ 
\partial^2_\tau \mu_0(j,0)=\int_{\mX_j} <P(A_j);A_j>_{T^\star \mX_j} 
d{\bf{x}}_j$, \\ 
which is  a pseudodifferential 
operator of order $0$ with principal symbol, the square matrix 
$p_0 (x,\xi )=(p_{0}^{ik}(x,\xi ))_{1\leq i, k\leq n-1}$ defined as follows. 
 In local coordinates, if $\Di \mh_j =\sum_{i,k} G_{i k}(x)dx_i dx_k ,$ then \\ 
$\Di \frac{|\mX_j|}{2}p_{0}^{ik}(x,\xi )\; =\; G^{ik}(x) \; -\; 
\sum_{\ell ,m} G^{im}(x)G^{\ell k}(x)\frac{\xi_m}{|\xi |} \frac{\xi_\ell }{|\xi |} \; ;\quad  
 (|\xi |^2=  \sum_{\ell ,m} G^{m\ell }(x)\xi_m \xi_\ell )\; ,$ \\ 
 so for any 
$\Di \zeta \in \R^{n-1}, \ \sum_{i,k} $ $\Di\frac{|\mX_j|}{2}p_{0}^{ik}(x,\xi )\zeta_i \zeta_k 
=\frac{2}{|\mX_j|} [ |\zeta |^2 -\frac{<\xi ; \zeta>^2}{|\xi |^2}] \geq 0;\ $
$\Di (<\xi ; \zeta>=\sum_{i,k}G^{ik}\xi_i \zeta_k ).$ \\
Thus  we get
$ \  
\partial^2_\tau \mu_0(j,0 ) =\int_{\mX_j}< P(A_j);A_j>_{T^\star \mX_j} d{\bf{x}}_j\;  >0\; \Carre $

\begin{lemma}\label{Poincare} For a one-form $A$ satisfying (\ref{hypA2}), there exists 
a constant $C_A>0$ such that, if $u$ is a function in 
$L^2(\mM )$ such that  
$du\in L^2(\mM )$ and 
\begin{equation}\label{coefFour} 
\forall j=1,\ldots , J,\quad \int_{\mX_j} u(x_j,y)d{\bf{x}}_j =0\; ,\quad \forall y\in ]a_j^2, +\infty[ \; , 
\end{equation} 
then 
$\forall \tau \in ]0,1],$ 
\begin{equation}\label{inequaPoincare} 
\| idu +\tau uA \|^{2}_{L^2(\mM )} \; \leq \; (1+\tau C_A) \| idu \|^{2}_{L^2(\mM )} + C_A 
\| u  \|^{2}_{L^2(\mM )} \; . 
\end{equation}  
\end{lemma} 
{\bf {Proof of Lemma \ref{Poincare}. }} First we remark that the inequality 
\begin{equation}\label{inequa1} 
| idu +\tau uA|^2 \leq (1+\rho ) |du|^2 +(1+\rho^{-1} ) | \tau uA|^2 
\end{equation} 
is satisfied for any $\rho >0.$ 

For $\rho =\tau   $ we get that there exists a constant $C_A^0>0$, depending only 
on $A/\mM_0,\ $   such that 
\begin{equation}\label{inequa2} 
\| idu +\tau uA \|^{2}_{L^2(\mM_0 )} \; \leq \; (1+\tau ) \| idu \|^{2}_{L^2(\mM_0 )} + \tau C^0_A 
\| u  \|^{2}_{L^2(\mM_0 )} \; . 
\end{equation}  
We get also for $\rho =\tau  $ that for any $j\in \{1, \ldots , J\} ,$ 
\begin{equation}\label{inequa3} 
\int_{a_j^2}^{+\infty} \| idu +\tau uA \|^{2}_{L^2(\mX_j )} y^{(2-n)\delta_j} dy \; \leq \; 
\end{equation}  
$$ \int_{a_j^2}^{+\infty}   \left ((1+\tau )  \| idu \|^{2}_{L^2(\mX_j )} + \tau C^j_A 
\| u  \|^{2}_{L^2(\mX_j )} \right )  y^{(2-n)\delta_j} dy  \; , $$ 
for some constant $C^j_A$  depending only 
on $A/X_j .$

But (\ref{coefFour}) implies that 
\begin{equation}\label{inequa4} 
 \| u  \|^{2}_{L^2(\mX_j )} \leq \frac{1}{\mu_1 (j,0)} \| idu \|^{2}_{L^2(\mX_j )} \; , 
\end{equation} 
with $(\mu_\ell (j,0))_{\ell \in \N } $ the sequence of eigenvalues of Laplace-Beltrami 
operator on $\mX_j,\ \mu_0 (j,0)=0 <\mu_1 (j,0)\leq \mu_2 (j,0)\leq \ldots  .$ 
So if (\ref{coefFour}) is satisfied then (\ref{inequa3}) and (\ref{inequa4}) 
imply that 
\begin{equation}\label{inequa5} 
 \| idu +\tau uA \|^{2}_{L^2(\mM_j )}  \; \leq \; 
   (1+\tau c^j_A)  \| idu \|^{2}_{L^2(\mM_j )}    \; ,  
\end{equation}  
for some constant $c^j_A$  depending only 
on $A/X_j .$

The existence of a constant $C_A >0$ satisfying the inequality (\ref{inequaPoincare}) follows from (\ref{inequa2}) 
and (\ref{inequa5}) for $j=1, \ldots J \; \Carre $

\begin{lemma}\label{LThD} 
When $A$ satisfies (\ref{hypA}),  (\ref{hypA2}) and 
Lemma \ref{FirstEige} , then as $\lambda \to +\infty ,$ 
 the following Weyl formula  is satisfied. 
\begin{equation}\label{LDe} 
N(\lambda, -\Delta_{(\lambda^{-\rho}A)} )\; =\; |\mM| \frac{\omega_n}{(2\pi )^n}   \lambda^{n/2} 
\; + \; \bf{O}(r_0(\lambda ) ) \; ,
\end{equation} 
with 
\begin{equation}\label{defRo}
 \rho \; =\; \left \{ \begin{array}{lll} 1/2, & \MR{if} & 2/n\leq \delta \leq 1 \\ 
(n\delta -1)/2 , & \MR{if} &    1/n < \delta < 2/n 
\end{array} \right . , 
\end{equation} 
$\Di \delta $ and $\omega_d$ are as in Theorem \ref{ThD}, 
and the function $ r_0(\lambda ) $ is the one defined by (\ref{LrL}) .

\end{lemma}
{\bf {Proof of Lemma \ref{LThD}. }} We follow the proof  
of Theorem \ref{ThD}. 

Since   $A$ satisfies Lemma \ref{FirstEige}, 
we have for $\lambda > >1$ large enough that  
$-\Delta_{(\lambda^{-\rho}A)} -(-\Delta_0)$ 
is in $\mM_0$  a partial differential operator of order $1$ with bounded coefficients, 
 so the part of the proof of  Theorem \ref{ThD} in $\mM_0$ remains valid for the estimate of  
$N(\lambda , -\Delta^{\mM_0,Z}_{(\lambda^{-\rho}A)}),\ (Z=D$ or $Z=N),$ because 
for any $\Di \Lambda >>1,\quad 
 N(\Lambda , -\Delta^{\mM_0,Z}_{0}+C(-\Delta^{\mM_0,Z}_{0})^{1/2} +C)  
\leq N(\Lambda , -\Delta^{\mM_0,Z}_{(\lambda^{-\rho}A)}) 
\leq N(\Lambda , -\Delta^{\mM_0,Z}_{0}-C(-\Delta^{\mM_0,Z}_{0})^{1/2} -C)$\\ 
and $\Di |N(\Lambda , -\Delta^{\mM_0,Z}_{0}\pm C(-\Delta^{\mM_0,Z}_{0})^{1/2} \pm C) 
- |\mM_0| \frac{\omega_n}{(2\pi )^n} \Lambda^{n/2} |\leq C \Lambda^{(n-1)/2} .$ 

For the part of the proof of Theorem \ref{ThD} in $\mM_j,\ 1\leq j,$  
we have also for any $\Lambda >>1,\quad 
N(\Lambda , -\Delta^{\mX_j}_{0}+C(-\Delta^{\mX_j}_{0})^{1/2} +C)  
\leq N(\Lambda , -\Delta^{\mX_j}_{(\lambda^{-\rho}A_j)}) 
\leq N(\Lambda , -\Delta^{\mX_j}_{0}-C(-\Delta^{\mX_j}_{0})^{1/2} -C)$\\ 
and $\Di |N(\Lambda , -\Delta^{\mX_j}_{0}\pm C(-\Delta^{\mX_j}_{0})^{1/2} \pm C) 
- |\mX_j| \frac{\omega_{n-1}}{(2\pi )^{n-1}} \Lambda^{(n-1)/2} |\leq C \Lambda^{(n-2)/2} .$ \\ 
But the crucial step of the proof of  Theorem \ref{ThD} 
is  Lemma \ref{Titch}, where we used, (with $\mu_\ell (j) $ to be replaced by $ \mu_\ell (j,1)$ in our new notations),  that
$$0<C\leq \mu_0(j)\leq \mu_\ell (j)\leq \mu_{\ell +1}(j)\quad \MR{and}\quad \lim_{\ell \to +\infty} \mu_\ell (j)=+\infty .$$ 
Here  in $\mM_j, (1\leq j),$ if $(\mu_\ell (j,\lambda ^{-\rho}) )_{\ell \in \N }$ denotes 
the increasing sequence of eigenvalues of 
$-\Delta_{\lambda^{-\rho}A_j}^{\mX_j}, $ 
we have 
$$C/\lambda^{2\rho} \leq \mu_0(j, \lambda ^{-\rho})
\quad {\rm and}\quad C\leq \mu_1(j,\lambda ^{-\rho})\leq \mu_{1+\ell}(j,\lambda ^{-\rho}) 
\leq \mu_{2+\ell}(j,\lambda ^{-\rho})$$ 
 with  $\Di  \lim_{\ell \to + \infty} 
 \mu_{\ell}(j,\lambda ^{-\rho})=+\infty \; .$\\ 
 More precisely  $\Di \lim_{\lambda \to +\infty} \mu_\ell (j,\lambda ^{-\rho}) = 
 \mu_\ell (j,0) $ and $0=\mu_0 (j,0) <\mu_{1+\ell} (j,0) $ for any $\ell \in \N .$ 
It follows that Lemma \ref{Titch} holds for any $\ell \in K_\lambda , \ell \neq 0$.
 So taking (\ref{NL1}) 
 into account, the proof  
of Theorem \ref{ThD} will remain valid if we can prove, (for $L^{D}_{j,0}$ as in Lemma  \ref{Titch}, excepted that   
 $\mu_0(j)$ is replaced by $\mu_0(j,\lambda ^{-\rho})$),  
 that 
$$N(\lambda ,L^{D}_{j,0} )\; =\; {\bf{O}}(r_0(\lambda ))\; .$$ 
This  can  easily be done as follows.

When $\delta_j=1, \ (\rho =1/2),$ it is easy to see that 
$$N(\lambda ,L^{D}_{j,0} )\; \leq \; N(\lambda +C , L^{D,\lambda} )\; \leq \; C\lambda^{1/2}\ln (\lambda ) \; ,$$ 
where $L^{D,\lambda}$ is the Dirichlet operator on $]0,+\infty [$ associated to 
$\Di \frac{C}{\lambda }e^{2t} - \partial_t^2\; .$ 
 
 When $0<\delta_j<1,$ by scaling we have that 
$$N(\lambda ,L^{D}_{j,0} )\; \leq \; N((\lambda +C)^{1+2\rho (1-\delta_j)} , L^D)\; \leq \; 
C\lambda^{(1+2\rho (1-\delta_j))/(2\delta_j)}\; ,$$ 
where $L^D$ is the Dirichlet operator on $]0,+\infty [$ associated to $\Di \frac{1}{C^2} t^{\frac{2\delta_j}{1-\delta_j}} - \partial_t^2\; .$  \\    
When $\Di 2/n \leq \delta <1,$ as $ 2/n \leq   \delta \leq \delta_j ,$  then \\ 
$\lambda^{(1+2\rho (1-\delta_j))/(2\delta_j)}=\lambda^{(2-\delta_j)/(2\delta_j)}
\leq \lambda^{(2-\delta )/(2\delta )} \leq \lambda^{(n-1 )/2}
= {\bf{O}}(r_0(\lambda )) .$ \\ 
When $\Di 1/n < \delta <2/n,$ as $\delta \leq \delta_j ,$ then \\ 
$\lambda^{(1+2\rho (1-\delta_j))/(2\delta_j)} \leq \lambda^{(1+2\rho (1-\delta ))/(2\delta )} 
= \lambda^{(n- (n\delta -1 ))/2}
={\bf{O}}(r_0(\lambda )) 
\;  \Carre $

To achieve the proof of Theorem \ref{Coro}, we take a one-form $A$ satisfying the assumptions 
of Lemma \ref{LThD}. 

We remark that any eigenfunction 
$u$ of the Laplace-Beltrami operator $-\Delta $ on $\mM $ associated to an eigenvalue
 in $\Di ]\inf \sp_\ess (-\Delta ) ,+\infty [ ,$  satisfies (\ref{coefFour}). 
So if $H_\lambda $ is the subspace of $L^2 (\mM )$ spanned 
by eigenfunctions of $-\Delta $ associated to eigenvalues in $]0, \lambda[, $ 
then, by (\ref{inequaPoincare}) of Lemma \ref{Poincare} with $\tau =1/\lambda ^\rho  ,$ 
with $\rho $ defined by (\ref{defRo}), 
we have 

$$
\forall u\in H_\lambda ,\quad \| idu +\frac{1}{\lambda ^\rho} uA\|^{2}_{L^2(\mM )} \; 
\leq \; (1+\frac{C_A}{\lambda ^\rho} ) \| du \|^{2}_{L^2(\mM )} +C_A \| u \|^{2}_{L^2(\mM )} $$
$$\quad\quad\quad\leq \; \left((1+\frac{C_A}{\lambda ^\rho} ) \lambda + C_A \right) \| u \|^{2}_{L^2(\mM )}\ .
$$
But if $(\lambda_j)_{j\in \N }$ is the non decreasing sequence of 
eigenvalues of $-\Delta_{(\lambda^{-\rho }A)},$ 
then by max-min principle one must have 
$$k < {\rm dim}(H_\lambda )\; \Rightarrow \; \lambda_k < 
(1+\frac{C_A}{\lambda ^\rho} ) \lambda + C_A \; ;$$ 
so 
\begin{equation}\label{compar2} 
{\rm dim}(H_\lambda ) \;  \leq \;  N\left((1+\frac{C_A}{\lambda ^\rho}) \lambda + C_A, -\Delta_{(\lambda^{-\rho }A)} \right)\; +\; 1
\; . 
\end{equation} 
The estimates (\ref{LDe}) and (\ref{compar2}) prove (\ref{Maj}), 
by noticing that $\lambda^{n/2}/\lambda^\rho ={\bf{O}}(r_0(\lambda ))\;  \Carre $

\noindent 
{\bf Acknowledgements} 

{\it {One of the authors is  grateful to Sylvain Gol\'enia  
for his useful comments.

{\it We also thank the referee for his useful and pertinent remarks.}


\begin{thebibliography}{}

\end{thebibliography}


\begin{thebibliography}{99}  


\bibitem[{Co-Le}]{Co-Le} E.A. Coddington, N. Levinson~: 
\newblock Theory of ordinary differential equation,
\newblock McGraw-Hill, New York, (1955), TMH Edition 1972, Fifteenth reprint 1996



\bibitem[{Col1}]{Col1} Y. Colin de Verdi\`ere~: 
\newblock Pseudo-laplaciens II,
\newblock Ann. Institut Fourier, 33 (2), (1983), p. 87-113. 





\bibitem[{Do-Li}]{Do-Li} H. Donnelly, P. Li~: 
\newblock Pure point spectrum and negative curvature for noncompact manifold,
\newblock Duke Math. J. 46, (1979), p. 497-503. 

\bibitem[{Don1}]{Don1} Harold Donnelly~: 
\newblock On the point spectrum for finite volume symmetric spaces of negative curvature,
\newblock Comm. in P.D.E., 6(9), (1981), p. 963-992. 


\bibitem[{Don2}]{Don2} Harold Donnelly~: 
\newblock On the essential spectrum of a complete riemannian manifold,
\newblock Topology, vol. 20, (1981), p. 1-14. 


\bibitem[{Go-Mo}]{Go-Mo} S. Gol\'enia, S. Moroianu~: 
\newblock Spectral Analysis of Magnetic Laplacians on 
Conformally Cusp Manifolds,
\newblock Ann. Henri Poincar\'e, 9, (2008), p. 131-179. 



\bibitem[{Hej}]{Hej} D. Hejhal~: 
\newblock The Selberg trace formula for $PSL(2, \bR ),$ II,
\newblock Lecture Notes in Math. 1001, Springer-verlag, Berlin, 1983. 



\bibitem[{Hel}]{Hel} B. Helffer~: 
\newblock  Effet d'Aharonov Bohm sur un \'etat born\'e de l'\'equation de Schr\"odinger, 
\newblock Commun. Math. Phys. 119, (1988), p. 315-329. 

\bibitem[{Hor1}]{Hor1} Lars H\"ormander~: 
\newblock The spectral function of an elliptic operator, 
\newblock Acta Math., 88, (1968), p. 341-370.


\bibitem[{Hor2}]{Hor2} Lars H\"ormander~: 
\newblock The Analysis of Linear P.D.O. IV ,
\newblock Springer-verlag, Berlin, 1985. 

\bibitem[{Ivr}]{Ivr} V. J. Ivrii~: 
\newblock Microlocal Analysis and Precise Spectral Asymptotics,
\newblock Springer-verlag, Berlin, 1998. 



\bibitem[{Mo-Tr}]{Mo-Tr3} A. Morame, F. Truc~: 
\newblock Eigenvalues of Laplacian with constant magnetic field 
on non-compact hyperbolic surfaces with finite area
 spectral asymptotics,
\newblock Lett. Math. Phys.,  vol. 97(2), (2011), p.203-211.




\bibitem[{Mul}]{Mul} W. M\"uller~: 
\newblock Weyl's law in the theory of automorphic forms, 
\newblock London Math. Soc. Lecture Notes, 354,(2008), p. 133-163. 

\bibitem[{Per}]{Per} A. Persson~: 
\newblock Bounds for the discrete part of the spectrum of a semibounded Scr\"odinger operator, 
\newblock Math. Scand.,  8,(1960), p. 143-153. 



\bibitem[{Sa-Va}]{Sa-Va} Y. Safarov, D. Vassiliev~: 
\newblock The Asymptotic distribution of Eigenvalues of Partial  Differential Operators,
\newblock  AMS Trans.155, 1996. 

\bibitem[{Shu}]{Shu} M. Shubin~: 
\newblock The essential Self-adjointness 
for Semi-bounded Magnetic Schr\"odinger operators on 
Non-compact Manifolds, 
\newblock J. Func. Anal., 186, (2001), p. 92-116. 



\bibitem[{Tit}]{Tit} E. C. Titchmarsh~: 
\newblock Eigenfunction Expansions associated
with Second-order Differential equations I, second edition,
\newblock  Oxford at the Clarendon Press, 1962.


\end{thebibliography}
\end{document}